\documentclass[11pt,a4paper]{article}

\pdfoutput=1
\usepackage{jheppub}

\usepackage{epsfig}
\usepackage{hyperref}
\usepackage{amssymb}
\usepackage{amsbsy}
\usepackage{amsmath}
\usepackage{url}

\newcommand{\rf}[1]{(\ref{#1})}
\newcommand{\beq}{\begin{equation}}
\newcommand{\eeq}{\end{equation}}
\newcommand{\be}{\begin{equation}}
\newcommand{\ee}{\end{equation}}
\newcommand{\bea}{\begin{eqnarray}}
\newcommand{\eea}{\end{eqnarray}}
\newcommand{\eq}[1]{Eq.~(\ref{#1})}
\newcommand{\non}{\nonumber \\*}
\newcommand{\ie}{{i.e.}\ }
\newcommand{\bv}{\begin{verbatim}}
\newcommand{\ev}{\end{verbatim}}

\newcommand{\vp}{\varphi}

\newcommand{\e}{\,\mbox{e}}
\renewcommand{\d}{{\rm d}}

\newcommand{\blambda}{\bar\lambda}
\newcommand{\brho}{\bar\rho}
\newcommand{\C}{\blambda}
\newcommand{\A}{A}

\newcommand{\half}{{\textstyle \frac 12}}

\newcommand{\bz}{{\bar z}}
\newcommand{\p}{\partial}
\newcommand{\bp}{\bar\partial}

\renewcommand{\b}{B}
\renewcommand{\c}{C}

\newcommand{\eps}{\varepsilon}
\newcommand{\om}{\omega}

\newcommand{\tr}{\mathrm{tr}}
\newcommand{\LA}{\left\langle}
\newcommand{\RA}{\right\rangle}

\def\fun#1#2{\lower3.6pt\vbox{\baselineskip0pt\lineskip.9pt
\ialign{$\mathsurround=0pt#1\hfil##\hfil$\crcr#2\crcr\sim\crcr}}}

\begin{document}



\title{Pauli-Villars' regularization of ghosts in path-integral string formulation}

\author{Yuri Makeenko}
\vspace*{2mm}
\affiliation{NRC ``Kurchatov Institute''\/-- ITEP,  Moscow\\
}
\emailAdd{makeenko@itep.ru} 


\abstract
{I consider Pauli-Villars' regulators for the ghosts in the path-integral string formulation
and show how they preserve conformal invariance.
I calculate the regulator contributions to the effective action and to the central charge and
demonstrate the consistency of the mean-field quantization of  the Nambu-Goto 
string in $2< d\leq26$. 
The higher-derivative corrections to the Liouville action are briefly considered  for
the Pauli-Villars and proper-time regularizations.}




\maketitle

 \section{Introduction}
 This Paper continues my previous Article~\cite{Mak18} 
 on the mean-field quantization of an effective string.
 The motivation for that was a global instability of the usual classical ground state
 of the Nambu-Goto string in the target-space dimension $d>2$. 
 On the contrary the mean-field ground state is perturbatively stable for $2<d\leq 26$ 
 both under global and under wavy local fluctuations, while the classical one is stable only for
 $d<2$.  The latter  describes a vast amount of the models of Statistical Mechanics
 and the former is associated  with the QCD string in $d=4$. 
 
 The idea of Ref.~\cite{Mak18} was
 first to deal with the mean-field approximation which sums up an infinite number of bubble
 diagrams of perturbation theory about the classical ground state 
 and then to consider a loop expansion about it. 
 Such an approach perfectly works, for instance, in the two-dimensional $O(N)$-symmetric
 sigma-model where the loop expansion has the meaning of the $1/N$-expansion.
The one-loop correction to the mean field was explicitly computed~\cite{Mak18} using
 the Pauli-Villars regularization for the target-space coordinate $X^\mu$.
 
 One of the most interesting results of Ref.~\cite{Mak18} is that the mean-field
 quantization of the Nambu-Goto string is apparently consistent for any $2<d\leq 26$ contrarily to
 the usual canonical quantization which is consistent only in $d=26$. 
  It was proposed that the usual central charge $(d-26)$, where $d$ comes from $X^\mu$
  and $-26$ comes from the ghosts, cancels
 in the mean-field approximation against
 the contribution from the Pauli-Villars regulators  which equals $(26-d)$.
 This is like for the noncritical Polyakov string where the consistency is linked to
 the presence of the Liouville field. However, only the Pauli-Villars 
 regulators for $X^{\mu}$ were considered and shown to give $-d$. 
 Those for the ghosts were not considered.
 The objective  of this Paper is to introduce Pauli-Villars' regulators for the ghosts and
 to show they to add $+26$ to the central charge.
 
 My motivation for writing this Paper has been also the recent interest~\cite{Mak21,Mak22,Mak22c,ST22} in
 higher-derivative actions of two-dimensional gravity. So
 one more goal of this Paper is to develop the technique for computing the higher-derivative corrections to the
 effective action, governing fluctuation of the metric, which emerges after the path integration over
 $X^\mu$, its regulators,
 ghosts and their regulators. For the Schwinger
 proper-time regularization 
a part of it is known 
 from the DeWitt-Seeley expansion~\cite{DeWitt,DeWitt2,Gil75}  
 of the heat kernel in the UV cutoff $\eps$.
 But this expansion applies to the path integrating over $X^\mu$ rather than to ghosts, 
 where only
 the leading order (the conformal anomaly) is known. Using the Pauli-Villars regularization for
 $X^\mu$ and  ghosts, I compute below the expansion of  both determinants
 applicable for computing higher-derivative terms in the effective action.
 
 
 The organization of the Paper is as follows. After a brief reminding of the setup in 
 Sect.~\ref{s:2}, I consider in Sect.~\ref{s:3} the Pauli-Villars regulators for the ghosts.
 These regulators are massive fields but still preserving conformal invariance.
 Then I compute the contribution from the ghost regulators to the effective action
 in Sect.~\ref{s:4} and to the central charge in Sect.~\ref{s:5}, demonstrating the consistency
 of the mean-field quantization in $2< d\leq26$.

\section{The setup\label{s:2}}   

Let us begin with reminding the Nambu-Goto action of the bosonic string which is the area
of the string worldsheet. It is highly nonlinear in $X^\mu$ but
can be made quadratic, introducing the Lagrange multiplier 
$\lambda^{ab}$ and an independent metric tensor $g_{ab}$, as
\be
S_{\rm NG}=K_0\int \sqrt{\det {(\p_a X\cdot \p_b X)}}=
K_0\int \Big [\sqrt{g} +\frac 12\lambda^{ab} (\p_a X\cdot \p_b X-g_{ab}) \Big],
\label{SNG}
\ee
where $K_0=1/2\pi\alpha'$ stands for the bare string tension.
We consider a closed string which wraps along the  compactified dimension
of circumference  $\beta$ and propagates through the distance $L\gg\beta$ which can  also be
compactified. The string worldsheet has thus topology of a cylinder or a torus.
There is no tachyon for such a string configuration
if $\beta$ is such to guarantee that  the classical energy of
the string dominates over the energy of zero-point fluctuations.

In the mean-field approximation the path integral over the (imaginary) Lagrange multiplier
$\lambda^{ab}$ has a saddle point at 
\be
\lambda^{ab}=\blambda \sqrt{g }\,g^{ab},
\ee
where $\blambda$ is constant for the proper choice of the worldsheet coordinates. 
Classically $\blambda=\blambda_{\rm cl}=1$
so the action~\rf{SNG} reduces to the action of the Polyakov string
\be
S=\frac{K_0}{2}\int \sqrt{g} g^{ab} \p_a X \cdot \p_b X
\label{SPol}
\ee
which is quadratic in $X^\mu$ that makes it easy to integrate it out
in the path integral. 
It is convenient to diagonalize $g_{ab}$,
choosing the conformal gauge where
$g_{ab}=\rho \delta_{ab}$, so that $\sqrt{g}=\rho$.
This procedure adds ghosts which are the same as for the Polyakov string formu\-lation.

The mean-field values of $\blambda$ and $\brho$ are calculated for both  a cylinder~\cite{AM15}
and  
a torus~\cite{AM21}.
The result is remarkable simple
\be
\blambda=\frac{1}2 +\frac{\Lambda^2}{2K_0} +
\sqrt{\frac 14\left(1 +\frac{\Lambda^2}{K_0}\right)^2 -\frac{d\Lambda^2}{2K_0}},
\label{newC} 
\ee
where $\Lambda$ is the UV cutoff,
and
\be
\brho= 
\frac \C {\sqrt{ \left(1 +\frac{\Lambda^2}{K_0}\right)^2 -\frac{2d\Lambda^2}{K_0}}} 
\rho_{\rm cl}
\label{newrho}
\ee
for $\beta \gg \Lambda^{-1}$. Here $\rho_{\rm cl}=1$ for the worldsheet parametrization.

Equation \rf{newC} is well-defined if the bare string tension 
\be
K_0>K_*=\left(d-1+\sqrt{d^2-2d} \right)\Lambda^2.
\label{K*}
\ee
At the critical value  $K_0=K_*$ the square root in \rf{newC} vanishes.
The classical ground state is recovered by Eqs.~\rf{newC}, \rf{newrho} as $K_0\to\infty$, 
while the expansion in $1/K_0$ 
makes sense of the semiclassical (perturbative) expansion about this vacuum.
The usual one-loop results are recovered to order $1/K_0$.
The value of $\blambda$ decreases with decreasing $K_0$ from the classical value
 $\blambda_{\rm cl}=1$ at $K_0=\infty$ 
to the {quantum} value
\be
\C_*=\frac 12\left(d -\sqrt{d^2-2d} \right)
\label{laq}
\ee
at $K_0=K_*$.

The metric \rf{newrho} becomes infinite when $K_0\to K_*$ given by \eq{K*}, which
is crucial for constructing the Lilliputian scaling limit~\cite{AM16a}.
Classically $\brho$  is simply the induced metric $\rho_{\rm cl}$
but in the mean-field approximation $\brho$ coincides with the averaged induced metric
$
\LA \partial_a X \cdot \partial_b X \RA=\brho \delta_{ab}
$,
where the average is understood in the sense of the path integral. 

To path integrate over $X^\mu$ we split 
$X^\mu=X^\mu_{\rm cl} +X^\mu_{\rm q}$ and perform  the Gaussian path
integral over $X^\mu_{\rm q}$ to
obtain the effective action governing fluctuations of $\lambda^{ab}$ and $g_{ab}$.
The corresponding determinant of the two-dimensional operator
\be
{\cal O}= -\frac 1\rho \p_a \lambda^{ab}\p_b
\label{defO}
\ee
is divergent %
 and is conveniently regularized by adding the Pauli-Villars regulators
\be
{S}_{\rm reg}=\frac {K_0}2 \int \left (\lambda^{ab} \p_a Y \cdot\p_b Y
 +M^2\sqrt{g} \, Y^2 \right),
\label{Sreg}
\ee
where every loop of the regulator field $Y^\mu$ brings the minus sign to compensate divergences
coming from $X^\mu$. Actually, we have to have~\cite{AM17c} two such regulators 
of mass squared $M^2$ with wrong statistics and one regulator of mass squared $2M^2$
with normal statistics to regularize all the divergences including the ones in tadpole
diagrams. Integrating out $X^\mu$ with its regulators
and the ghost with their regulators 
and minimizing the emergent effective action with respect to  $\lambda^{ab}$ and $\rho$,
we arrive at Eqs.~\rf{newC} and \rf{newrho} with%
\footnote{To be exact $M^2 \to M^2/\blambda$ in this and below formulas.\label{f:1}}
\be
\Lambda^2=\frac{M^2}{2\pi}\log2
\label{LaM}
\ee
for the described Pauli-Villars regularization.

\section{Ghost regulators as conformal fields\label{s:3}}

The action of massive Grassmannian ghost fields%
\footnote{I capitalize the letters denoting the ghosts to emphasize they are massive.} 
$\c^a$ and (traceless) $\b_{ab}$ reads~\cite{Dia89}
\be
S_{\rm gh}=  \int \sqrt{g} \left(g^{ik}\b_{ij}\nabla_k \c^j +  m \epsilon_{ij} \c^i \c^j
-2  m g^{ik} \epsilon^{jl} \b_{ij}\b_{kl}\right),
\ee
where $\epsilon_{ij}$ is the covariant Levi-Civita symbol, or
\be
{\cal S}_{\rm gh}=  \int \left(\b_{zz}\bp \c^z +\b_{\bz\bz}\p \c^\bz +m \e^{2\vp} \c^\bz \c^z
 +m \e^{-\vp}\b_{\bz\bz} \b_{zz}\right)
 \label{gS}
\ee
in the conformal gauge. The classical equations of motion then read
\bea
\bp \c^z-m \e^{-\vp}\b_{\bz\bz}=0, \qquad \bp \b_{zz}- m \e^{2\vp} \c^\bz=0.  
\label{gcem}
\eea

For the nonvanishing propagators we have
\be
\LA \c^z \b_{\om\om} \RA =\LA \b_{zz} \c^\om \RA=
-4 \p G_{2m}(z-\om),  \quad  
\LA \c^z \c^{\bar \om} \RA=\LA  \b_{zz}\b_{\bar\om\bar \om} \RA=2 m G_{2m}(z-\om),~
\label{gprop}
\ee
where 
\be
G_m(z)=\frac1{2\pi}K_0 (m\sqrt{z\bz})
\ee
with $K_0 (m\sqrt{z\bz})$ being the modified Bessel function
is the massive propagator, reproducing  usual
\be
\LA c^z b_{\om\om} \RA =\LA b_{zz} c^\om \RA=\frac 1{\pi(z-\om)}
\ee
as $m\to0$.

The $T_{zz}$ component of the energy-momentum tensor is not changed for $m\neq 0$
\be
T^{(\rm gh)}_{zz} = \pi \Big[\c^z \p \b_{zz} + 2(\p \c^z) \b_{zz} \Big]
\label{Tzzgh}
\ee
but now the $T_{z\bz}$ component
\be
T^{(\rm gh)}_{z\bz} = m \e^{-\vp} \b_{\bz\bz} \b_{zz}  -2 m \e^{2\vp} \c^\bz \c^z  
\ee
does not vanish. 

Nevertheless, the total energy-momentum tensor which is the sum of the
one for $X^\mu$ plus its regulators and the one for the ghosts plus their regulators is traceless thanks to
the classical equation of motion for $\vp$. This is a general property because
\be
\hat g^{ab} \frac{\delta {\cal S}[g]}{\delta \hat g^{ab} }= -
\frac{\delta {\cal S}[g]}{\delta \vp }
\label{tra}
\ee
for
\be
g_{ab}=\hat g_{ab} \e^\vp.
\label{confog}
\ee
The left-hand side of \eq{tra} represents the trace of the energy-momentum tensor
while the right-hand side represents the classical equation of motion for $\vp$.
This is analogous to the tracelessness of  the ``improved'' energy-momentum tensor 
in two dimensions~\cite{DJ95,Jac05} which is always
traceless  thanks to the classical equation of motion. The conformal Ward identities are thus usual
in spite of the presence of the massive regulator fields.

The same consideration applies in the mean-field approximation, 
where the vanishing of the right-hand side of
\eq{tra} is precisely the minimization condition that determines the mean field. We thus expect the mean-field approximation to enjoy conformal 
invariance in a full analogy with the classical theory.
In Sect.~\ref{s:5} I explicitly demonstrate this by the calculation of the central charge.

\section{Ghost contribution to effective action\label{s:4}}

\begin{figure}
\centerline{\includegraphics[width=8cm]{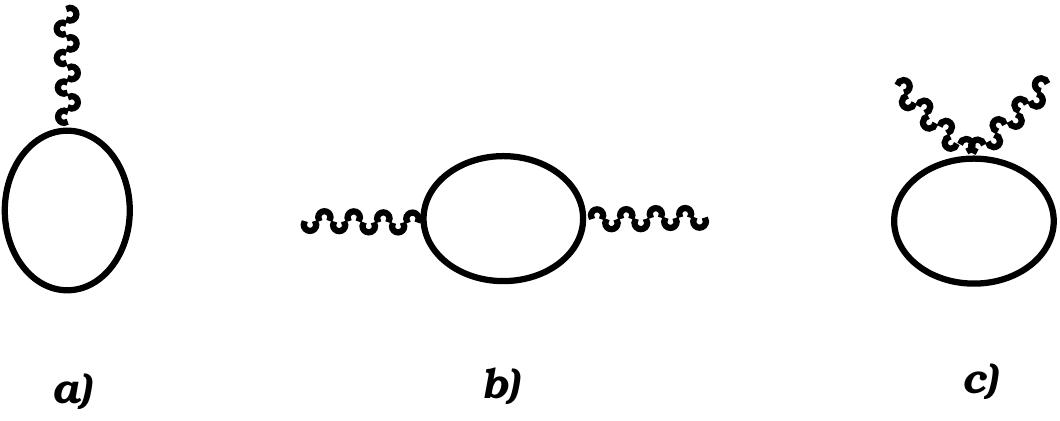}} 
\caption{One-loop diagrams for the effective action of $ \vp$.}
\label{fi:1g}
\end{figure}

Given~\rf{gS}, \rf{gprop} we can compute the contribution of the ghosts to the effective action,
represented at the one-loop order by the diagrams in Fig.~\ref{fi:1g}. The interaction
vertices come from the expansion of \rf{gS} in $\vp$ and change chirality of the ghost
regulators. To regularize the divergences we apply the Pauli-Villars regularization of the ghosts,
adding to the usual (massless) ghosts with Fermi statistics the massive fields discussed
in the previous section. These additional massive ``ghosts'' should obey Bose statistics to cancel
the divergences coming from the usual ghosts. This is not yet the whole story because
actually one needs like in Ref.~\cite{AM17c} two of such regulators plus one regulator with 
Fermi statistics and the mass $\sqrt{2}m$ to cancel all the divergences. This regularization
of the ghost determinant preserves conformal invariance, as argued in the previous section,
so one does not expect any logs to appear.

Let us apply the Pauli-Villars regularization of the ghost determinant to the diagrams
in Fig.~\ref{fi:1g}.
It is easy to compute the tadpole diagram in Fig.~\ref{fi:1g}$a$, where we have either the loop
of the $c$-ghost regulator with the coefficient 2 or  the loop of the $b$-ghost regulator
with the coefficient -1. Its contribution to the effective action is
\be
\hbox{Fig.~\ref{fi:1g}$a$}=\left(2 \int \frac{\d^2 k}{(2\pi)^2} \frac {M^2}{(k^2+M^2)} -
\int \frac{\d^2 k}{(2\pi)^2} \frac {2M^2}{(k^2+2M^2)}\right) \vp  = 
\vp\frac {M^2}{2\pi}\log2 ,
\ee
where we have introduced%
\footnote{If the masses of the $\b$- and $\c$-ghosts 
are different, $m_b$ and $m_c$ respectivety, then the formulas
below still hold with $M^2=4 m_b m_c$.}
 $M=2m$ to comply with the Pauli-Villars regularization of $X^\mu$ (cf.~\eq{LaM}).
The diagram in Fig.~\ref{fi:1g}$c$ together  with the divergent%
\footnote{Here and below ``divergent'' means divergent as $M\to\infty$.}
part of the diagram in Fig.~\ref{fi:1g}$b$ gives
\be
\hbox{Fig.~\ref{fi:1g}$c$} +\hbox{Fig.~\ref{fi:1g}$b$}_{\rm div}= 
\frac{\vp^2}2\frac {M^2}{2\pi}\log2 
\ee
as is required for $\e^\vp$.

In the calculation of 
the finite part of the diagram in Fig.~\ref{fi:1g}$b$ we can restrict ourselves with only
one regulator because the contribution of the two others cancels in the term $p^2$. 
But we keep below all
three regulators to correctly compute the terms of the order $p^4 M^{-2}$
and higher. We then write first the contribution
from only one regulator and then repeat for all three.

The result is conveniently expressed via the integral
\be
p^2 G\Big(\frac {p^2}{M^2}\Big)  \equiv
- 24 \pi \int \frac{\d^2 k}{(2\pi)^2} 
\frac {M^4}{(k^2+M^2)[(p-k)^2+M^2]}
=-12
\frac{M^4 \,{\rm arctanh}\, 
\frac{\sqrt{p^2(p^2+4M^2)}}{p^2+2M^2}}{ \sqrt{p^2(p^2+4M^2)}}
\label{defG}
\ee
 which is already familiar from
 the matter sector (\ie coming from $X^\mu$ and its regulators), where
it reads~\cite{Mak18}
\bea
\hbox{Fig.~\ref{fi:1g}$b$}\big|_{\rm fin} &=& -\frac d{96\pi} p^2
\left(2 G \Big(\frac {p^2}{M^2}\Big)- G \Big(\frac {p^2}{2M^2}\Big) \right)|\vp|^2 \non
&=&
-\frac {d}{96 \pi} p^2\left(1-\frac{3 p^2}{10 M^2} 
+{\cal O}\big(M^{-4}\big) \right) |\vp|^2.
\label{1mat}
\eea

For the ghosts
the diagram in Fig.~\ref{fi:1g}$b$ involves the loop with either the $\vp \bar \b \b$
 times  $\vp \bar \c \c$ vertices or the $(\vp \bar \b \b)^2$ plus  $(\vp \bar \c \c)^2$ vertices.
 The former has the coefficient $- 2\cdot 2=-4$ and the latter has the one
 $1+2^2=5$. All together
we have in the finite part of the diagram in Fig.~\ref{fi:1g}$b$ for
the contribution from one regulator
\bea
&&\frac12 \int \frac{\d^2 k}{(2\pi)^2}
\left(4\frac {M^2k^a(p-k)_a}{(k^2+M^2)[(p-k)^2+M^2]} 
-5 \frac {M^4}{(k^2+M^2)[(p-k)^2+M^2]
} \right.\non &&\hspace*{2.1cm}  \left.+\frac {4M^2k^2}{(k^2+M^2)^2}
\right) |\vp|^2 =\frac {1}{48 \pi}
\left(\frac{2p^2}{M^2}-1 \right) p^2G\Big(\frac {p^2}{M^2}\Big) |\vp|^2
\label{Igh}
\eea
and for the contribution from all three
\bea
\hbox{Fig.~\ref{fi:1g}$b$}\big|_{\rm fin} &=& \frac1{48\pi}p^2 \left[
2 \left(\frac{2p^2}{M^2}-1 \right) G\Big(\frac {p^2}{M^2}\Big) -\left(\frac{p^2}{M^2}-1 \right)
 G\Big(\frac {p^2}{2M^2}\Big)\right]
 |\vp|^2
\non &=&\frac{1}{96\pi} p^2
\left(26-\frac{33 p^2}{5 M^2} +{\cal O}\big(M^{-4}\big)
\right) |\vp|^2 .
\label{1gh}
\eea

The sum of  \rf{1mat} and \rf{1gh} reproduces as $M\to\infty$ the usual shift of 
$d$ by $-26$ in the 
effective action thanks to the ghosts. I also kept there the ${\cal O}\big(M^{-2}\big)$
corrections which contribute to the curvature-squared term in the effective action as 
will be momentarily discussed.

A little bonus from the above computation is the exact formula for the effective action 
to quadratic order in $\vp$
\be
{\cal S}_{\rm eff} = \frac 1{96\pi h}\int \frac{\d^2 p}{(2\pi)^2} p^2 F\Big(\frac {p^2}{M^2}\Big) |\vp|^2
\label{SF}
\ee
with
\bea
F \Big(\frac {p^2}{M^2}\Big) &=&
(2-d)\left(2 G\Big(\frac {p^2}{M^2}\Big)- G \Big(\frac {p^2}{2M^2}\Big) \right)-
\frac{2p^2}{M^2}\left(4 G\Big(\frac {p^2}{M^2}\Big)-G\Big(\frac {p^2}{2M^2}\Big) \right) 
\non &=&
26-d+\frac{3d p^2}{10 M^2} -\frac{33 p^2}{5 M^2}
+{\cal O}\big(M^{-4}\big) 
\label{defF}
\eea
and the function $G$ defined in \eq{defG}.
The function $F( M^{-2}p^2)$ is positive for $d\leq26$ and decreases from the value
 $26\!-\!d$ to $0$
with increasing $p^2$.
It is not universal (except for the first term of the expansion) 
and depends on the regularization applied to 
compute the determinants.  

As is already pointed out, 
the higher-order terms of the function $F(p^2/M^2)$ in \eq{SF} are 
regularization-dependent and \eq{defF} refers to the Pauli-Villars regularization
(this $F$ to be called $F^{({\rm PV})}$). 
As shown in Appendix~\ref{appA}, the coefficients of the Taylor expansion of
 $F^{({\rm PV})}(p^2/M^2)$
are linked  to those of 
$F^{({\rm S})}(p^2 \eps)$ for the 
proper-time regularization as
\be
F^{({\rm PV})}_n=  2\left( 1-2^{-n}  \right) (n-1)! F^{({\rm S})}_n .
\ee
Equation~\rf{SF} with $F= F^{({\rm S})}(p^2 \eps)$ determines the effective action for the 
proper-time regularization.

Covariantizing \rf{SF} we write
\be
{\cal S} _{\rm eff} = -\frac 1{96\pi h}\int  \sqrt{g}  \vp \Delta F( -M^{-2}\Delta) \vp
\label{cSF}
\ee
which is the higher-derivative action of the type discussed in Refs.~\cite{Mak21,ST22}.
For the Polyakov string there are no other contributions to this order in $h$.
For the Nambu-Goto string an additional (nonlocal) higher-derivative term emerges~\cite{Mak21} 
after the path integration over the Lagrange multiplier
$\lambda^{ab}$. In the gauge~\rf{confog} 
with vanishing scalar curvature $\hat R$ of the background metric $\hat g_{ab}$ it reads
\be
\delta {\cal S} _{\rm eff} \propto 
M^{-2}h^{-1} \int \sqrt{g} g^{ab}
\p_a \vp \p_b \vp \Delta \vp +{\cal O}\big(M^{-4}\big).
\label{addit}
\ee
The coefficient is calculable with the given technique and  the result will be presented
elsewhere.

\section{Central charge in mean-field approximation\label{s:5}}

To compute the central charge, we use the total 
energy-momentum tensor $T_{zz} $ of the Nambu-Goto string which is the sum  
 of the ones for the field $X^\mu$ 
\be
T_{z z} ^{X}= 2\pi K_0 \left[\lambda ^{z\bz}(\p X)^2  
+\lambda^{\bz\bz} \p X \cdot \bp X \right], 
\label{TX}
\ee
its regulator $Y^\mu$
\be
T_{z z} ^{Y}= 2\pi K_0 \left[\lambda ^{z\bz}(\p Y)^2  
+\lambda^{\bz\bz} \p Y \cdot \bp Y \right], 
\label{TY}
\ee
ghosts and ghost regulators given by \eq{Tzzgh}. 
In the mean-field approximation we can substitute $\lambda^{ab}$ by its mean value
$\bar \lambda^{z\bz}=\blambda$, $\bar \lambda^{\bz\bz}=0$, so the energy-momentum
tensors~\rf{TX} and \rf{TY} reproduce those for the Polyakov string except for now $\blambda\neq1$.

\begin{figure}
\centerline{\includegraphics[width=7.cm]{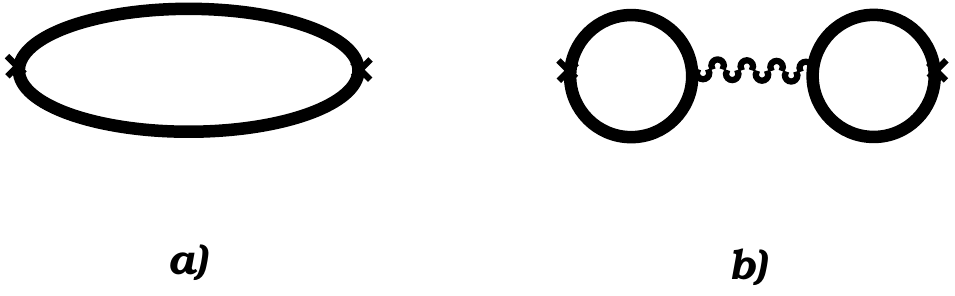} }
\caption{Diagrams contributing to the correlator $\LA T_{zz}(z) T_{zz}(0) \RA$  in the mean-field approximation.}
\label{mean-F2}
\end{figure}

The diagrams contributing to the correlator
$ 
\LA T_{zz}(z) T_{zz}(0) \RA
$ 
in the mean-field approximation are depicted in
 Fig.~\ref{mean-F2},
where the solid line corresponds either to 
the field $X^\mu$ (and its regulators) or to the ghosts (and their regulators), while the wavy line
corresponds to $\vp$ with the propagator
\be
\LA \vp(z) \vp(0) \RA =-\frac {12 h}{(26-d) } \log(z\bar z).
\label{rhorho}
\ee
For the diagram in Fig.~\ref{mean-F2}$a$ the regulator contributions to the central charge
vanish as $M\to\infty$ and 
it gives (omitting $h^{-1}$) the usual $d$ for $X^\mu$ and $-26$ for the ghosts~\cite{FMS86}.

An additional contribution comes 
from the diagram in Fig.~\ref{mean-F2}$b$ which is usually associated with the next order 
of the perturbative expansion about the classical vacuum
because it has two loops, but in the mean-field
approximation it has to be considered together with the diagram in  Fig.~\ref{mean-F2}$a$ 
since it of the same order in $h$: $h^{-1}\, h \;h^{-1}\sim h^{-1}$.
Each of the two closed loops of the Pauli-Villars regulators of $X^\mu$ in the diagram in  Fig.~\ref{mean-F2}$b$ contributes
in momentum space (see footnote~\ref{f:1})
\be
2\pi d \int \frac{\d^2 k}{(2\pi)^2} \, 
\frac{k_z(p_z-k_z) M^2}{(k^2+M^2)[(k-p)^2+M^2]}= \frac{d}{12} p_z^2.
\label{intzz}
\ee
The analogous contribution from the regulators of ghosts reads
\be
-4\pi  \int \frac{\d^2 k}{(2\pi)^2} \, 
\frac{5k_z(p_z-k_z) M^2 +4 k_z^2 M^2}{(k^2+M^2)[(k-p)^2+M^2]}= -\frac{13}{6} p_z^2.
\label{gintzz}
\ee
In the sum of \rf{intzz} and \rf{gintzz}
the ghost contribution remarkably shifts the matter contribution $d$ by $-26$ just like in the 
effective action.

Multiplying the contribution of the two loops
in the diagram in Fig.~\ref{mean-F2}$b$ by the fourth derivative of the propagator, we obtain
\be
\hbox{Fig.~\ref{mean-F2}$b$}=
\frac {(d-26)}{12} \frac {12} {(26-d)} \frac {(d-26)}{12} \frac 6{z^4}=\frac {(26-d)}{2 z^4}. 
\ee

The sum of the diagrams in Fig.~\ref{mean-F2}$a$ and $b$ thus gives for the central charge
\be
c = (d-26) +(-d+26)=0 
\label{c=0}
\ee
which illustrates the consistency of the mean-field quantization.%
\footnote{The vanishing of the total central charge guarantees the vanishing of
the Weyl anomaly for the gravitational background $\hat g_{ab}$ in \eq{confog}.}
It is applicable as long as the effective action is stable which means $2< d\leq26$.
Thus in the mean-field approximation the situation with the central charge of
the Nambu-Goto and Polyakov strings are the same although the ground states are
different. It can be viewed as a consequence of the background  independence.

It is worth noting the relation of the consideration in the previous paragraph with the 
original Polyakov formulation of the noncritical string~\cite{Pol81} where the path integration
over $X^\mu$ and the ghosts results in the Liouville action for the field $\vp$.
Equation~\rf{c=0} then represents the  usual compensation of the
central charge of $X^\mu$ plus the ghosts that equals $d\!-\!26$
by the one of the Liouville field which equals $26\!-\!d$.

For the  Nambu-Goto string we have additionally the path integration over the Lagrange multiplier $\lambda^{ab}$ 
which yields~\cite{Mak21} the additional term~\rf{addit} in the effective action for $\vp$.
It has a quartic derivative and is therefore negligible for smooth classical $\vp$ but  nevertheless it revives in quantum
computations as a result of doing uncertainties $M^{-2}\times M^2$.
As shown in Refs.~\cite{Mak22,Mak22c} this higher-derivative term gives an additional contribution
to the central charge of $\vp$ at one loop and accordingly changes
the string susceptibility at the one-loop order, telling the Nambu-Goto and Polyakov strings apart.

Yet another argument showing that the Nambu-Goto and Polyakov
strings may be not equivalent at one loop is the deviation of excited states from the Alvarez-Arvis
string spectrum discovered in \cite{DFG12,AK13,Hel14} for an open outstretched Nambu-Goto
string. It shows up again at the one-loop order. It is hard to understand this deviation for the Polyakov string
which was one of my original motivations for studying conformal properties of the Nambu-Goto string.

\subsection*{Acknowledgement}

I am grateful to Arkady Tseytlin for useful correspondence.
This work was supported by the Russian Science Foundation (Grant No.20-12-00195).


\appendix

\section{Proper-time versus Pauli-Villars regularizations\label{appA}}

For the Polyakov string the path integration over the target-space coordinates
$X^\mu$ and the ghosts was performed    
in 1980's by the DeWitt-Seeley expansion
using Schwinger's proper-time regulatization. The determinant
of an elliptic operator  $ {\cal O}$ is then regularized by
\be
\tr \log ({\cal O})\big|_{\rm reg}= -\int _\eps^\infty \frac {\d \tau}{\tau}\tr \e^{-\tau {\cal O}}.
\ee

 For the Pauli-Villars regularization 
the determinants are conveniently regularized  by the ratio
of massless to massive determinants as~\cite{AM17c}
\be
\det({\cal O})\big|_{\rm reg}\equiv\frac{\det({\cal O})\det({\cal O}+2M^2)}{\det({\cal O}+M^2)^2},
\label{newR}
\ee
so that
\be
 \tr\log({\cal O})\big|_{\rm reg}= -
\int_{0}^\infty \frac{\d \tau}{\tau} \,\tr \e^{-\tau{\cal O}}
\left(1-\e^{-\tau M^2}\right)^2
\label{PV22}
\ee
is convergent. Here $M\to\infty$ is the regulator mass which is related to $\eps$ by
\be\frac 1\eps
= {M^2}\log 4.
\label{M2}
\ee
We have added in \rf{newR} the ratio of the
determinants for the masses $\sqrt{2}M$ and $M$ to cancel the
logarithmic divergence at small $\tau$, because   the DeWitt-Seeley 
expansion 
\be
\LA \omega \Big| \e^{-\tau {\cal O} }
\Big| \omega \RA =\sum _{n=0}^\infty A_n (\om)\tau^{n-1}
\label{sse}
\ee
starts with the term  
$1/\tau$ in two dimensions. More regulators are required 
in higher dimensions.

Just like the original (massless) determinant
the massive determinants in \eq{newR} can be represented as path integrals 
\be
\det \left( {\cal O} + M^2 \right)^{d}
=\int {\cal D} \bar Y^\mu {\cal D}  Y^\mu
\e^{-{K_0}\int  \sqrt{g}\left( \bar Y \cdot  {\cal O} Y + 
M^2 \, \bar Y\cdot Y \right)}
\ee
over two fields  $Y^\mu$ and  $\bar Y^\mu$ of mass squared $M^2$ with ghost statistics 
and one field $Z^\mu$  of mass squared $2M^2$
with normal statistics
\be
\det \left( {\cal O} + 2M^2 \right)^{-d/2}
=\int {\cal D} Z^\mu 
\e^{-\frac{K_0}2\int  \sqrt{g}\left( Z \cdot  {\cal O} Z + 
2M^2   Z^2 \right)}.
\ee

I shall briefly mention several advantages of the Pauli-Villars regularization over the 
proper-time regularization:
\begin{enumerate}
\vspace{-2mm}
\item Noether's currents are derivable in the regularized case.\vspace{-2.7mm}
\item The regulator mass term induces the interaction between the regulators and $\vp$ which
can be treated perturbatively by  Feynman's diagrammatic technique.
\vspace{-2.7mm}
\item A nonperturbative  Gel'fand-Yaglom method can be applied in symmetric cases.
\vspace{-2.7mm}
\item The Pauli-Villars regulators preserve conformal invariance as is discussed above.
\end{enumerate}

\vspace{-1mm}
For the proper-time regularization the variation of the effective action is expressed via 
the DeWitt-Seeley coefficients $A_n$'s as
\be
\frac{\delta {\cal S}_{\rm eff}[\vp]}{\delta \vp(\om)} = 
\e^{\vp(\om)} \sum _{n=0}^\infty A_n (\om)\eps^{n-1}.
\label{sse1}
\ee
For the Pauli-Villars regularization an additional factor emerges from the integral over $\tau$
in \eq{PV22} 
\be
\int_0^\infty \d\tau \, \tau^{n-1} \frac \partial {\partial \tau} \left( 1-\e^{-\tau M^2} \right)^2 =2\left( 1-2^{-n}  \right) (n-1)!
\label{iint}
\ee
and we obtain
\be
\frac{\delta {\cal S}_{\rm eff}[\vp]}{\delta \vp(\om)} = 
\e^{\vp(\om)} \left[ M^2 \log 4 +\sum _{n=1}^\infty 
2\left( 1-2^{-n}  \right) (n-1)!\,A_n(\om) M^{2-2n} \right].
\label{sse2}
\ee
The addional factor in \eq{sse2} equals 1 for $n=1$ and 3/2 for $n=2$. This extra 3/2 is 
what a doctor ordered to reproduce the known value of $A_2$~\cite{DeWitt} 
for the heat-kernel expansion  from \rf{1mat} derived for the Pauli-Villars regularization.

The same additional factors emerge also for the ghost contribution to the effective action while
the derivation is now a little bit more complicated.
The ghost determinant reads
$
\det ( P_1^\dagger P_1 )^{1/2}
$
with~\cite{Pol81}
\be
(P_1^\dagger P_1)_{ac} = -\nabla^b \left( g_{ac} \nabla_b +g_{bc} \nabla_a - g_{ab} \nabla_c \right) =\left(- \nabla^2 -\half R
\right) g_{ac} .
\ee
The variation of the ghost determinant involves also the conjugate operator
\be
( P_1 P_1^\dagger)_{ac} =\left( -\nabla^2+ R \right) g_{ac} .
\ee
according to  (cf.\ \cite{FTs81,DOP82,Alv83}) 
\be
\frac{\delta\, \tr\log (P_1^\dagger P_1)}{\delta \vp(\om)} =\int _0^\infty \d \tau \left(
-2 \LA\om\Big | \e^{-\tau P_1^\dagger P_1} \Big|\om\RA +
\LA\om\Big | \e^{-\tau P_1 P_1^\dagger } \Big|\om\RA \right) 
\frac \partial {\partial \tau} \left( 1-\e^{-\tau M^2} \right)^2.
\ee
In the analogous formula for the proper-time regularization the derivative $\p /\p\tau$ acts
on the Heaviside step function $\theta(\tau-\eps)$ producing $\delta(\tau-\eps)$. Thus the additional factor 
for the Pauli-Villars regularization is given again by
\eq{iint}.

\end{document}